\documentclass[graybox]{svmult}

\usepackage{mathptmx}       
\usepackage{helvet}         
\usepackage{courier}        
\usepackage{type1cm}        
\usepackage{makeidx}         
\usepackage{graphicx}        
\usepackage{multicol}        
\usepackage[bottom]{footmisc}

\makeindex             

\begin{document}

\title*{Nuclei in Strongly Magnetised Neutron Star Crusts}
\author{Rana Nandi and Debades Bandyopadhyay}
\authorrunning{Magnetised Neutron star crusts} 
\institute{Rana Nandi, Astroparticle Physics and Cosmology Division, Saha 
Institute of Nuclear Physics, 1/AF Bidhannagar, Kolkata-700064, India \email{rana.nandi@saha.ac.in}
\and Debades Bandyopadhyay, Astroparticle Physics and Cosmology Division and
Centre for Astroparticle Physics, Saha Institute of Nuclear Physics, 1/AF 
Bidhannagar, Kolkata-700064, India \email{debades.bandyopadhyay@saha.ac.in}}
%
%
\maketitle

\abstract{We discuss the ground state properties of matter in outer and inner 
crusts of neutron stars under the influence of strong magnetic fields. In 
particular, we demonstrate the effects of Landau quantization of electrons on 
compositions of neutron star crusts. First we
revisit the sequence of nuclei and the equation of 
state of the outer crust adopting the Baym, Pethick and Sutherland (BPS) model 
in the presence of strong magnetic fields and most recent versions of the 
theoretical and experimental nuclear mass tables. Next we deal with nuclei in 
the inner crust. Nuclei which are arranged in a lattice, are immersed in a 
nucleonic gas as well as a uniform background of electrons in the inner crust. 
The Wigner-Seitz approximation is adopted in this calculation and each lattice
volume is replaced by a spherical cell. 
The coexistence of two phases of nuclear matter - liquid and gas, is 
considered in this case. We obtain the equilibrium nucleus corresponding to each
baryon density by minimizing the free energy of the cell. 
We perform this calculation using Skyrme nucleon-nucleon interaction with 
different parameter sets. We find nuclei 
with larger mass and charge numbers in the inner crust in the presence of 
strong magnetic fields than those of the zero field case for all 
nucleon-nucleon interactions considered here. However, SLy4 interaction has 
dramatic effects on the proton fraction as well as masses and charges of 
nuclei. This may be attributed to the behaviour of symmetry energy with 
density in the sub-saturation density regime. Further we discuss the 
implications of our results to shear mode oscillations of magnetars.}

\section{Introduction}
\label{sec:1}
Neutron star crust is a possible site where neutron rich heavy nuclei might 
reside. Extreme physical conditions exist at the crust of a neutron star. The 
temperature is $\sim 10^{10}$ K and the density varies from $10^4$ - $10^{14}$ 
g/cm$^3$
there. Recently, it was observed that certain neutron stars called magnetars 
had surface magnetic fields $\sim 10^{15}$ G. The internal fields could be 
several times 
higher than the surface fields of magnetars. Soft gamma repeaters (SGRs) are 
suitable candidates for magnetars \cite{dun,thom,watt1}. Giant flares were 
observed from SGRs in several cases. Those giant flare events are thought to be
the results of star quakes in magnetars. This might be attributed to the 
tremendous magnetic stress due to the evolving magnetic field leading to 
cracks in the crust. Quasi-periodic oscillations discovered in three giant 
flares are the evidences of torsional shear mode oscillations in magnetar 
crusts.

Such strong magnetic fields in magnetars are expected to influence charged 
particles such as electrons in the crust through Landau quantization. The 
effects of strong magnetic fields on dense matter in neutron star interior were
studied earlier \cite{Deb,pal}. It was also noted that atoms, molecules became 
more bound in a magnetic field \cite{Lai}.
In this article, we discuss the effects of strongly quantising magnetic 
fields on compositions and equation of state of the ground state matter in 
neutron star crusts and its connection to torsional shear mode oscillations.

We organise the article in the following way. Neutron star crusts in 
strong magnetic fields are described in sections 2, 3 and 4. Torsional shear 
mode oscillations of magnetars are discussed in section 5. Finally, we 
summarise in section 6.

\section{Crusts in Strong Magnetic Fields}
\label{sec:2}
We investigate compositions and equations of state (EoS) of outer and 
inner crusts in strong magnetic fields. Nucleons are bound in nuclei in the 
outer crust. Nuclei are immersed in a uniform background of electron gas which 
becomes relativistic beyond $10^7$ g/cm$^3$. Neutrons start to drip out of 
nuclei at higher densities. This is the beginning of the inner crust. In this 
case, nuclei are embedded both in electron and neutron gases. Magnetic fields 
may influence the ground state properties of crusts  either through magnetic 
field and nuclear magnetic moment interaction or through Landau quantisation of
electrons.  
In a magnetic field $\sim 10^{17}$ G, magnetic field and nuclear
magnetic moment interaction would not produce any significant change.
However such a strong magnetic field is expected
to influence charged particles such as electrons in the crust through Landau
quantization. Our main focus is to study the effects of Landau quantisation on
the ground state properties of neutron star crusts. Later we discuss shear
mode frequencies using our results of magnetised neutron star crusts. 

\subsection{Landau Quantisation of Electrons}
\label{subsec:2}
We consider electrons are noninteracting and placed under strongly quantising
magnetic fields. In the 
presence of a magnetic field, the motion of electrons is quantized in the 
plane perpendicular to the field. We do not consider Landau 
quantisation of protons because magnetic fields in question in this calculation
are below the critical field for protons. However, protons in nuclei would be 
influenced by a magnetic field through the charge neutrality condition. 
We take the magnetic field ($\overrightarrow{B}$) along 
Z-direction and assume that it is uniform throughout the inner crust.
If the field strength exceeds a critical value 
$B_c=m_e^2/e\simeq 4.414\times 10^{13}$G, then electrons become relativistic
\cite{Lai}. 
The energy eigenvalue of relativistic electrons in a quantizing magnetic field 
is given by
\begin{equation}
 E_e(\nu,p_z)=\left[p_z^2+m_e^2 + 2 e B \nu \right]^{1/2} ~,
\end{equation}
where $p_z$ is the Z-component of momentum, $\nu$ is 
the Landau quantum number. The Fermi momentum of electrons, $p_{F_{e,\nu}}$, is
obtained from the electron chemical potential in a magnetic field
\begin{equation}
p_{F_{e,\nu}}= \left[{\mu_e}^2-m_e^2 - 2 e B \nu \right]^{1/2}~.
\label{mu}
\end{equation}

The number density of electrons in a magnetic field is calculated as
\begin{equation}
 n_e=\frac{eB}{2\pi^2}\sum_{\nu=0}^{\nu_{max}}g_{\nu}p_{F_{e,\nu}}~,
\end{equation} 
where the spin degeneracy is 
$g_\nu = 1$ for the lowest Landau level ($\nu=0$) and $g_\nu = 2$ for all
other levels.

The maximum Landau quantum number ($\nu_{max}$) is obtained from
\begin{equation}
 \nu_{max}=\frac{{\mu_e}^2-m_e^2}{2eB}~.
\end{equation}
The energy density of electrons is,
\begin{eqnarray}
\varepsilon_e=
\frac{eB}{4\pi^2} \sum_{\nu=0}^{\nu_{max}}g_{\nu} 
\left(p_{F_{e,\nu}}\mu_e + (m_e^2 + 2 e B \nu) 
\ln \frac{p_{F_{e,\nu}}+\mu_e}{\sqrt{(m_e^2 + 2 e B \nu)}}\right)~.
\label{en}
\end{eqnarray}

Similarly the pressure of the electron gas is determined by
\begin{eqnarray}
P_e=
\frac{eB}{4\pi^2} \sum_{\nu=0}^{\nu_{max}}g_{\nu} 
\left(p_{F_{e,\nu}}\mu_e - (m_e^2 + 2 e B \nu) 
\ln \frac{p_{F_{e,\nu}}+\mu_e}{\sqrt{(m_e^2 + 2 e B \nu)}}\right)~.
\label{pr}
\end{eqnarray}

\begin{figure}[th]
\centerline{\includegraphics[width=6cm]{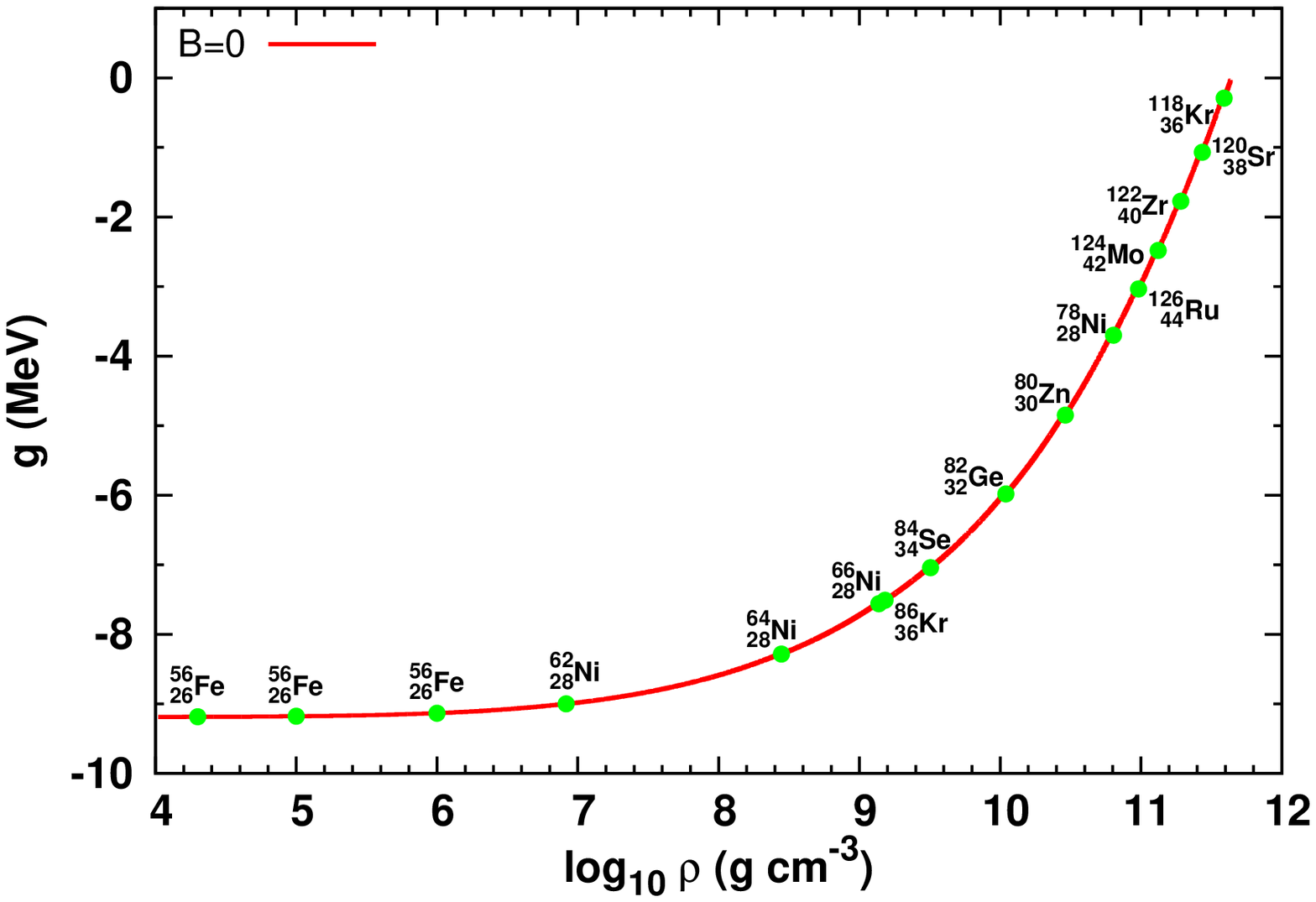}\includegraphics[width=6cm]{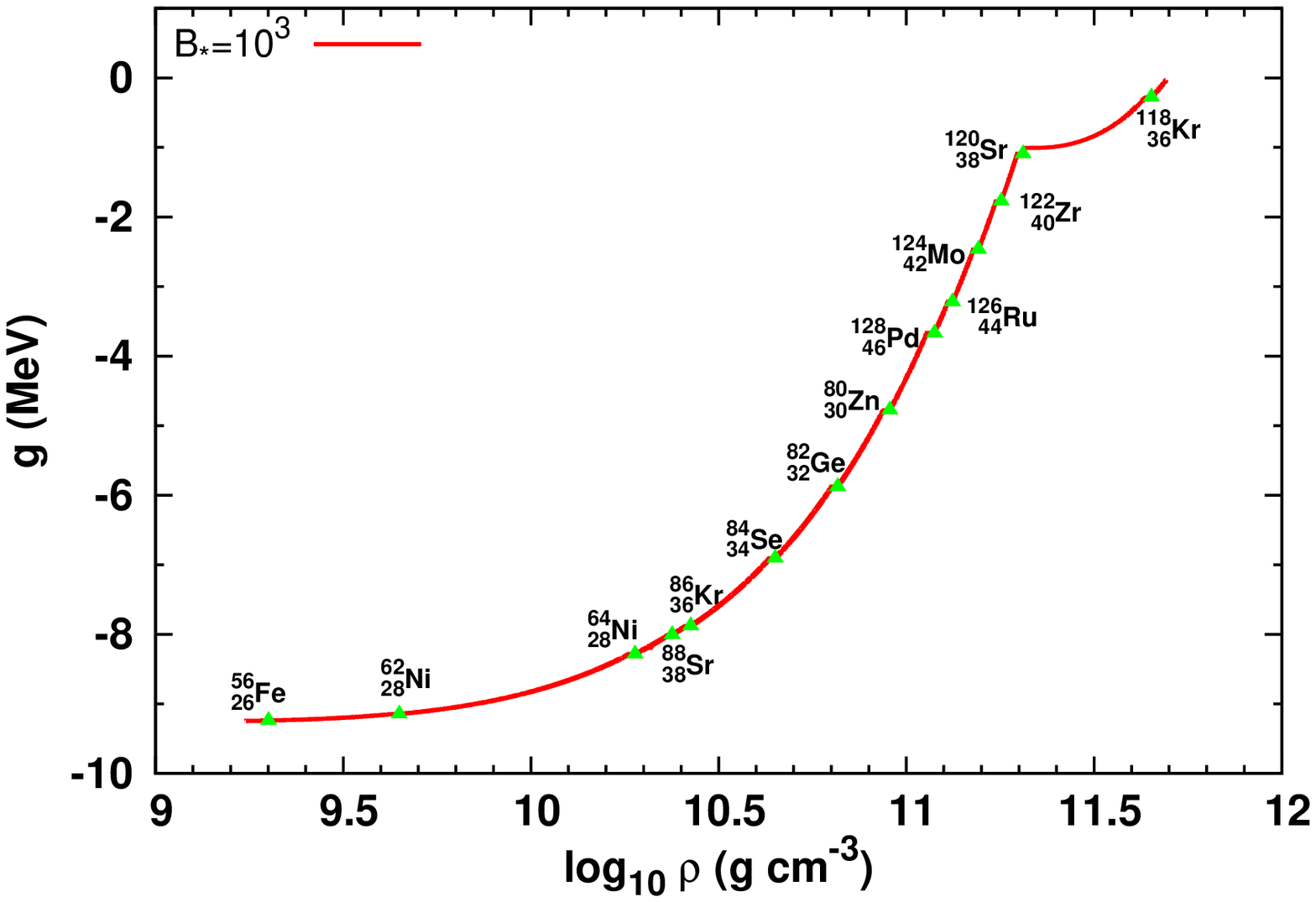}}
\caption{Gibbs free energy per nucleon is plotted with mass density for 
zero magnetic field (left panel) and $B_{*}=10^3$ (right panel). Equilibrium 
nuclei are shown with solid symbols in both panels.}
\label{fig1}       
\end{figure}

\section{Magnetic BPS Model of Outer Crust Revisited}
\label{sec:3}
We describe the BPS model in the presence of strong magnetic fields 
$B\sim 10^{16}$G to determine the sequence of equilibrium nuclei and 
the equation state of the outer crust \cite{lai91,rana1}. Nuclei are arranged 
in a bcc lattice in the outer crust. Here we adopt the Wigner-Seitz (WS) 
approximation and replace each lattice volume by a spherical cell which 
contains one nucleus at the center. Further each cell is to be charge neutral 
such that equal numbers of protons and electrons are present there. The 
Coulomb interaction among cells is neglected. An equilibrium nucleus
(A, Z) at a given pressure P is obtained by minimising the Gibbs free energy 
per nucleon with respect to A and Z. 
In this calculation, we modify the magnetic BPS model including the finite size
effect in the lattice energy and adopting recent experimental and theoretical 
mass tables. 
The total energy density of the system is given by
\begin{equation}
 E_{tot} = n_N (W_N+ W_L)+ \varepsilon_e~.
\end{equation}
The energy of the nucleus (including rest mass energy of nucleons) is
\begin{equation}
W_N = m_n (A - Z) + m_p Z - bA~,
\end{equation}
where $n_N$ is the number density of nuclei, $b$ is the binding energy per 
nucleon.
Experimental nuclear masses are obtained from the atomic mass table 
compiled by Audi, Wapstra and Thibault \cite{audi03}. For the rest of 
nuclei we use the theoretical extrapolation of M\"{o}ller et al.
\cite{moller95}. $W_L$ is the lattice energy of the cell and is given by
\begin{equation}
  W_L= -\frac{9}{10}\frac{Z^2e^2}{r_C}\left(1-\frac{5}{9} 
\left(\frac{r_N}{r_C}\right)^2\right)~.
\end{equation}
Here $r_C$ is the cell radius and $r_N\simeq r_0A^{1/3}$ ($r_0\simeq$1.16 fm) 
is the nuclear radius. The first term in $W_L$ is the lattice energy for point 
nuclei and the second term is the correction due to the finite size of the 
nucleus (assuming a uniform proton charge distribution in the nucleus). Further
$\varepsilon_e$ is the electron energy density as given by Eq.(\ref{en}) 
and P is the total pressure of the system given by
\begin{equation}
 P=P_e+\frac{1}{3}W_Ln_N~,
\end{equation}
where $P_e$ is the pressure of electron gas in a magnetic field as given by 
Eq.(\ref{pr}). 

The Gibbs free energy per nucleon is 
\begin{eqnarray}
g = \frac{E_{tot} + P}{n} = \frac{W_N + 4/3 W_L + Z \mu_e}{A}~,
\end{eqnarray}
where $n$ is the total baryon number density.

At a fixed pressure $P$, we minimise $g$ varying $A$ and $Z$ of a nucleus.
The sequence of equilibrium nuclei and their corresponding free
energies are shown in Fig.\ref{fig1}. Here we define $B_{*} = B/B_c$. The left 
panel shows results for $B=0$ 
and the right panel corresponds to $B_{*}=10^3$. It is evident from the 
figure that some nuclei disappear and new nuclei appear under the influence of
strong magnetic fields. It is attributed to the phase space modification of
electrons due to Landau quantisation which enhances the electron number density
\cite{rana1}.   

\section{Inner Crust in Quantizing Magnetic Fields}

Now we describe the ground properties of matter of inner crusts in presence of
strong magnetic fields using the Thomas-Fermi (TF) model at zero temperature.
Inner crust nuclei are immersed in a nucleonic gas as well as a uniform 
background of electrons. Furthermore, nuclei are arranged in a bcc lattice. As 
in the case of outer crust, we again adopt the Wigner-Seitz (WS) approximation 
in this calculation. Here each cell is taken to be charge neutral and the 
Coulomb interaction between cells is neglected. Electrons are uniformly 
distributed within a cell. The system is in $\beta$-equilibrium. We assume that
the system is placed in a uniform magnetic field. Though electrons are directly 
affected by strongly quantizing magnetic fields, protons in the cell are 
influenced through the charge neutrality condition \cite{rana2}. The 
interaction of nuclear magnetic moment with the field is not considered because
it is negligible in a magnetic field below $10^{18}$ G \cite{brod}.

The spherical cell in the WS approximation does not define a nucleus. We 
exploit the prescription of Bonche, Levit and Vautherin \cite{bon1,bon2} to 
subtract the gas part from the cell and obtain the nucleus. It was shown that 
the TF 
formalism at finite temperature generated two solutions \cite{sur} - one for 
the nucleus plus neutron gas and the other representing the neutron gas. The 
nucleus is obtained as the difference of two solutions. This formalism is 
adopted in our calculation at zero temperature as described below.

The thermodynamic potentials for nucleus plus gas (NG) and only gas (G) phases
are defined as \cite{bon1,bon2} 
\begin{equation}
\Omega = {\cal{F}} - \sum_{q=n,p} \mu_q n_q~,
\end{equation} 
where $\cal{F}$, $\mu_q$ and $n_q$ are the free energy density, baryon 
chemical potential and number density, respectively. The nucleus plus gas 
solution coincides with the gas solution at large distance i.e. 
$\Omega_{NG} = \Omega_{G}$. The free energy which is a function of 
baryon number density and proton fraction ($Y_p$), is defined as
\cite{rana2}
\begin{equation}
{\cal{F}}(n_q,Y_p) = \int [{\cal{H}} + \varepsilon_c + \varepsilon_e] d{\bf r}~.
\end{equation}

The nuclear energy density is calculated using the Skyrme nucleon-nucleon 
interaction and it is given by \cite{kri,bra,sto} 
\begin{eqnarray}
{\cal H}(r)&=&\frac{\hbar^2}{2m_n^*}\tau_n 
+\frac{\hbar^2}{2m_p^*}\tau_p+
\frac{1}{2}t_0\left[\left(1+\frac{x_0}{2}\right) n^2-\left(x_0+
\frac{1}{2}\right)\left(n^2_n+n^2_p\right)\right]\nonumber\\
&&-\frac{1}{16}\left[t_2\left(1+\frac{x_2}{2}\right)
-3t_1\left(1+\frac{x_1}{2}\right)\right](\nabla n)^2\nonumber\\
&&-\frac{1}{16}\left[3t_1\left(x_1+\frac{1}{2}\right)
+t_2\left(x_2+\frac{1}{2}
\right)\right]\left[(\nabla n_n)^2+(\nabla n_p)^2\right]\nonumber\\
&&+\frac{1}{12}t_3 n^\alpha\left[\left(1+\frac{x_3}{2}\right) n^2
-\left(x_3+\frac{1}{2}\right)\left(n_n^2+ n_p^2\right)\right]~,
\end{eqnarray} 
and the effective nucleon mass
\begin{eqnarray}
\frac{m}{m_q^*(r)}&=&1+\frac{m}{2\hbar^2}\left\{\left[t_1\left(1+
\frac{x_1}{2}\right)
+t_2\left(1+\frac{x_2}{2}\right)\right]\right. n \nonumber\\
&& +\left.\left[t_2\left(x_2+\frac{1}{2}\right)
-t_1\left(x_1+\frac{1}{2}
\right)\right] n_q\right\}~,
\end{eqnarray}
where total baryon density is $n = n_n + n_p$.

The direct parts of Coulomb energy densities for the nucleus plus gas and gas
phases follow from \cite{rana2,sil}
\begin{eqnarray}
\varepsilon^{NG}_c (r) &=&\frac{1}{2} (n^{NG}_p(r)-n_e) 
\int \frac{e^2}{\mid{\bf r}-{\bf r^{\prime}}\mid} 
(n^{NG}_p(r')-n_e)d{\bf r'}\nonumber\\
\varepsilon^{G}_c (r) &=&\frac{1}{2} (n^{G}_p(r)-n_e) 
\int \frac{e^2}{\mid{\bf r}-{\bf r'}\mid} 
(n^{G}_p(r')-n_e)d{\bf r'}\nonumber\\
&& +n^{N}_p(r)
\int \frac{e^2}{\mid{\bf r}-{\bf r^{\prime}}\mid} 
(n^{G}_p(r')-n_e)d{\bf r'}~,
\end{eqnarray}
where $n_p^{NG}$ and $n_p^G$ are proton densities in two respective phases. 
The exchange parts of coulomb energy densities are small and neglected in this
calculation.

The average electron chemical potential in a magnetic field given by 
Eq.({\ref{mu}}) is modified to \cite{rana2}
\begin{equation}
\mu_e = \left[p_{F_{e,\nu}}(\nu)^2+ m_e^2 + 2 eB \nu \right]^{1/2} - <V^c(r)>~,
\end{equation}
where $<V^{c}(r)>$ denotes the average single particle Coulomb potential and 
for both phases it is given by
\begin{equation}
 V^c(r) = \int\left[n^{NG}_p(r')-n_e\right]\frac{e^2}{\mid{{\bf r}
-{\bf r'}}\mid} d{\bf r'}~.
\end{equation}

The density profiles of neutrons and protons with or without magnetic fields are
obtained by minimising the thermodynamic potential in the TF approximation
\begin{eqnarray}
\frac{\delta \Omega_{NG}}{\delta n^{NG}_q}&=&0~,\nonumber\\ 
\frac{\delta \Omega_{G}}{\delta n^{G}_q}&=&0~,
\end{eqnarray}
with the condition of number conservation of each species from
\begin{eqnarray}
Z_{cell}&=&\int n_p^{NG}(r)d{\bf r}~,\nonumber\\
N_{cell}&=&\int n_n^{NG}(r)d{\bf r}~,
\end{eqnarray}
where $N_{cell}$ and $Z_{cell}$ are neutron and proton numbers in the cell, 
respectively.

We obtain the mass number $A = N +Z$ and atomic number using the subtraction
procedure as
\begin{eqnarray}
Z&=&\int \left[n_p^{NG}(r)- n_p^G(r)\right]d{\bf r}~,\nonumber\\
N&=&\int \left[n_n^{NG}(r)- n_n^G(r)\right]d{\bf r}~.
\end{eqnarray}

\begin{figure}[th]
\centerline{\includegraphics[width=6cm]{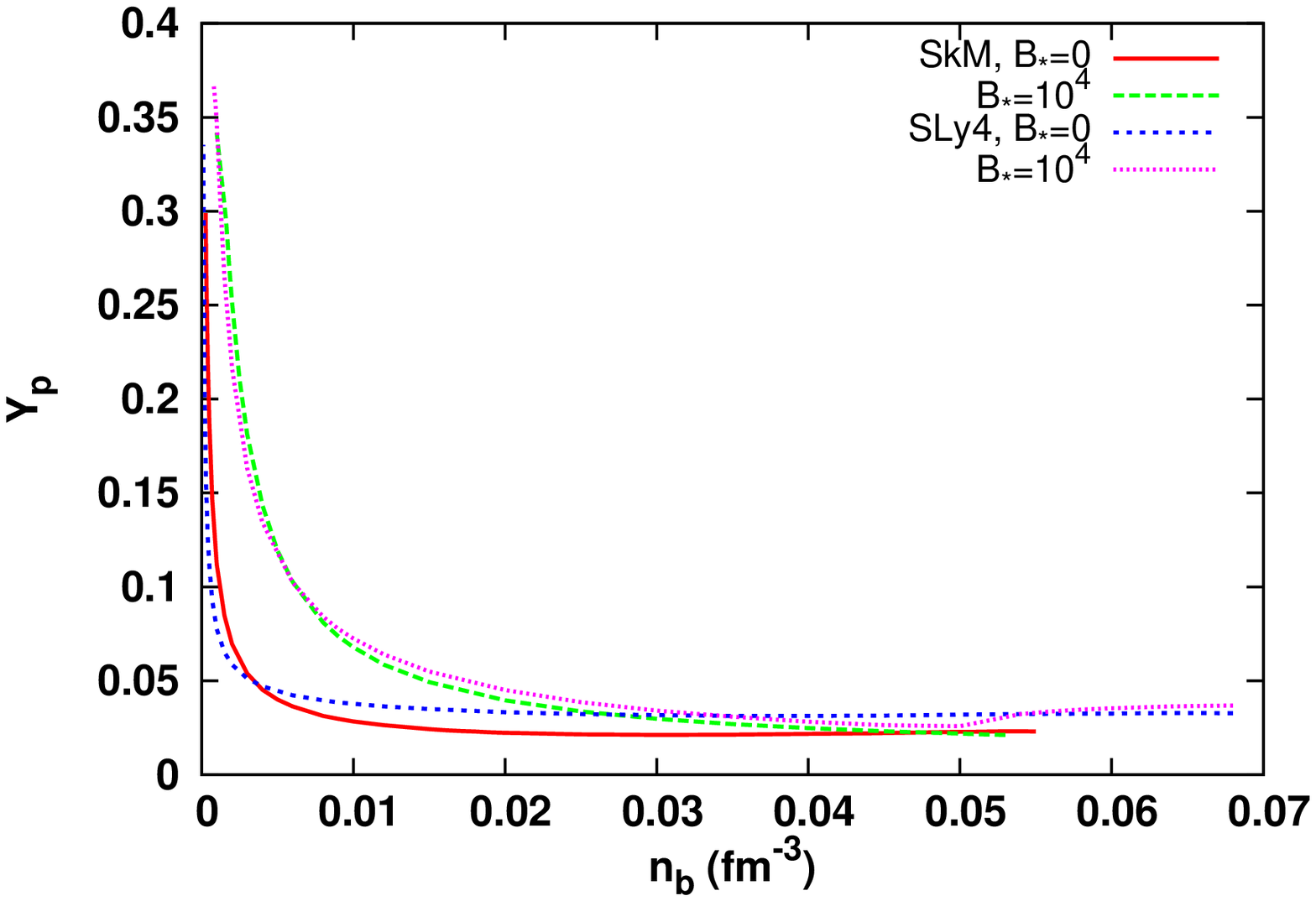}\includegraphics[width=6cm]{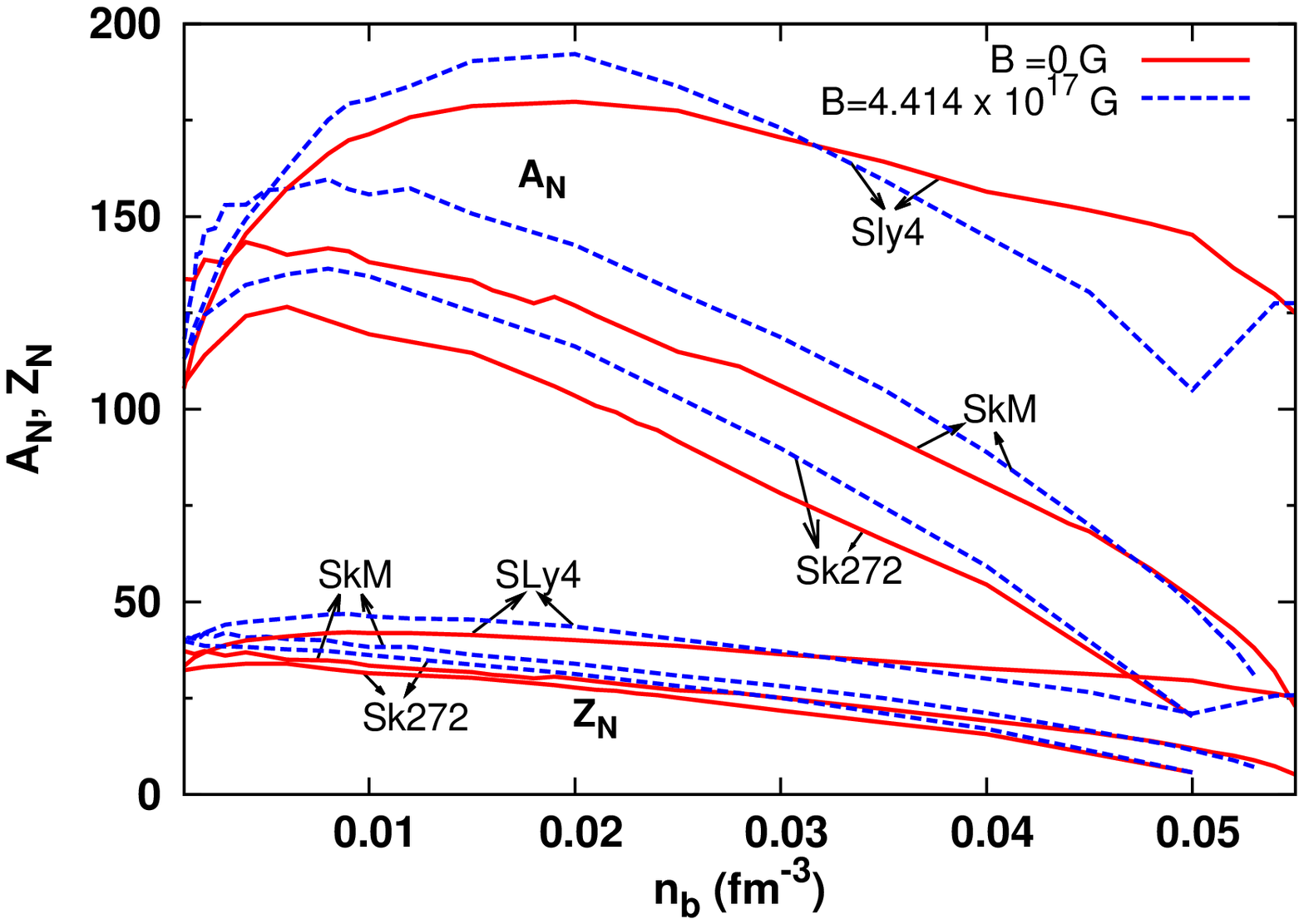}}
\caption{Proton fraction (left panel) and mass and atomic numbers of 
equilibrium nuclei (right panel) are plotted with average baryon density for 
different magnetic field strengths and Skyrme interaction parameter sets.}
\label{fig2}       
\end{figure}

Here we again obtain the equilibrium nucleus at each density by minimising the
free energy of the nuclear cluster in the cell along with charge neutrality 
and $\beta$-equilibrium conditions \cite{rana2}. In the left panel of 
Fig.\ref{fig2}, proton fraction is shown for $B=0$ and $B_{*} = 10^4$. 
Protons are influenced by the Landau quantisation of electrons through charge
neutrality condition. At lower densities, only the zeroth Landau level is
populated by electrons whereas a few Landau levels are populated above density
0.005 fm$^{-3}$ for $B_{*} = 10^4$ i.e. 4.414$\times 10^{17}$ G. This is 
reflected in the proton fraction which rises hugely at lower densities and 
approaches to the zero field case at higher densities. Further we estimate the
effects of different parameter sets of Skyrme interaction on the proton 
fraction. It is noted that the SLy4 set \cite{chab} results in higher proton 
fraction due to the stiffer density dependence of the symmetry energy at 
sub-saturation densities than that of the SkM set. 

We exhibit mass and atomic numbers of equilibrium nuclei after subtraction of 
free neutrons as a function of average baryon density in the right panel of
Fig.\ref{fig2}. Results are obtained for $B=0$ and $B_{*}=10^4$. Besides
SkM and SLy4 parameter sets, we also exploit Sk272 \cite{bij} parameter set for
this calculation. In all three cases, mass and atomic numbers are higher than 
zero field cases as long as only the zeroth Landau level is populated. However,
the situation is changed at higher densities when electrons jump from the
zeroth Landau level to the first level. This leads to jumps in mass and atomic
numbers in nuclei as noted for the SLy4 set. Further, the variation of 
parameters
for nucleon-nucleon interaction affects mass and atomic numbers of nuclei as
it is evident from the figure. We also note that the free energy of the ground 
state matter in strong magnetic fields is reduced and becomes more bound 
compared with the field free case.
 
\section{Shear Mode Oscillations in Magnetars}

Giant x-ray flares caused by the tremendous magnetic stress on the crust of 
magnetars were observed in several cases. Star quakes associated with these 
giant flares excite seismic oscillations. Quasi-periodic oscillations (QPOs) 
were found in the decaying tail of giant flares from SGR 1900+14 and 
SGR 1806-20. Those QPOs were identified as shear mode oscillations of magnetar 
crusts \cite{watt1,watt2}. Frequencies of the observed QPOs ranged from 18 Hz to
1800 Hz.

Shear mode frequencies are sensitive to the shear modulus of neutron star crust.
The shear modulus is again strongly dependent on the composition of neutron 
star crust. It might be possible to constrain the properties of neutron star 
crusts by studying the observed frequencies of QPOs. Torsional shear mode 
oscillations were investigated both in Newtonian gravity \cite{piro,mcd} and 
general relativity \cite{kip,sota1,sota2}. In both cases, it was assumed that 
the magnetised crust was decoupled from the fluid core.

Here we describe the calculation of shear mode frequencies adopting the model
of Sotani et al. \cite{sota1}. In this case, we study torsional shear 
oscillations of spherical and non-rotating relativistic stellar models. The 
metric used here has the form,
\begin{equation}
ds^2 = - e^{2\Phi} dt^2 + e^{2\Lambda} dr^2 + r^2 \left( d{\theta}^2 + sin^2{\theta} d{\phi}^2 \right)~.
\end{equation}

The equilibrium models are obtained by solving Tolman-Oppenheimer-Volkoff
equation. Next the equilibrium star is assumed to be endowed with a strong 
dipole magnetic 
field \cite{sota1}. The deformation in the equilibrium star for magnetic fields 
$\sim 10^{16}$ G is neglected. Torsional shear modes are the results of 
material velocity oscillations. These modes are incompressible and do not 
result in 
density perturbation in equilibrium stars. Consequently, this leads to 
negligible metric perturbations and justifies the use of the relativistic 
Cowling approximation \cite{sota1}. The relevant perturbed matter quantity for
shear modes is the $\phi$-component of the perturbed four velocity 
$\partial {u^{\phi}}$ \cite{sota1}
\begin{equation}
\partial {u^{\phi}} = e^{-\phi} \partial_t {\cal{Y}}(t,r) {\frac{1}{sin{\theta}}}{\partial_{\theta}} P_l(cos{\theta})~,
\end{equation}
where $\partial_t$ and $\partial_{\theta}$ correspond to partial derivatives
with respect to time and $\theta$, respectively, $P_l(cos{\theta})$ is 
the Legendre polynomial of order $\ell$ and ${\cal{Y}}(t,r)$ is the angular 
displacement of the matter. The perturbation equation is obtained
from the linearised equation of motion. Finally, we estimate eigenfrequencies
by solving two first order differential equations Eq.(69) and (70) of Sotani et
al. \cite{sota1}.

Now we study the dependence of shear mode frequencies on the compositions
of magnetised crusts which are already described in sections 3 and 4. 
Earlier calculations were performed with non-magnetised crusts 
\cite{watt1,sota1,sota2,stei}. One 
important input for the shear mode calculation is the knowledge of shear 
modulus of the magnetised crust. Here we adopt the expression of shear modulus 
as given by \cite{ichi,stroh} 
\begin{equation}
\mu = 0.1194 \frac{n_i (Ze)^2}{a}~,
\label{shr}
\end{equation}
where $a = 3/(4 \pi n_i)$, $Z$ is the atomic number of a nucleus and $n_i$ is
the ion density. This zero temperature form of the shear modulus was 
obtained by assuming a bcc lattice and performing directional averages 
\cite{han}. Later the dependence of the shear modulus on temperature was 
investigated with Monte Carlo sampling technique by Strohmayer et al. 
\cite{stroh}. However we use the zero temperature shear modulus of 
Eq.({\ref{shr}}) in this calculation.
\begin{figure}[th]
\centerline{\includegraphics[width=6cm]{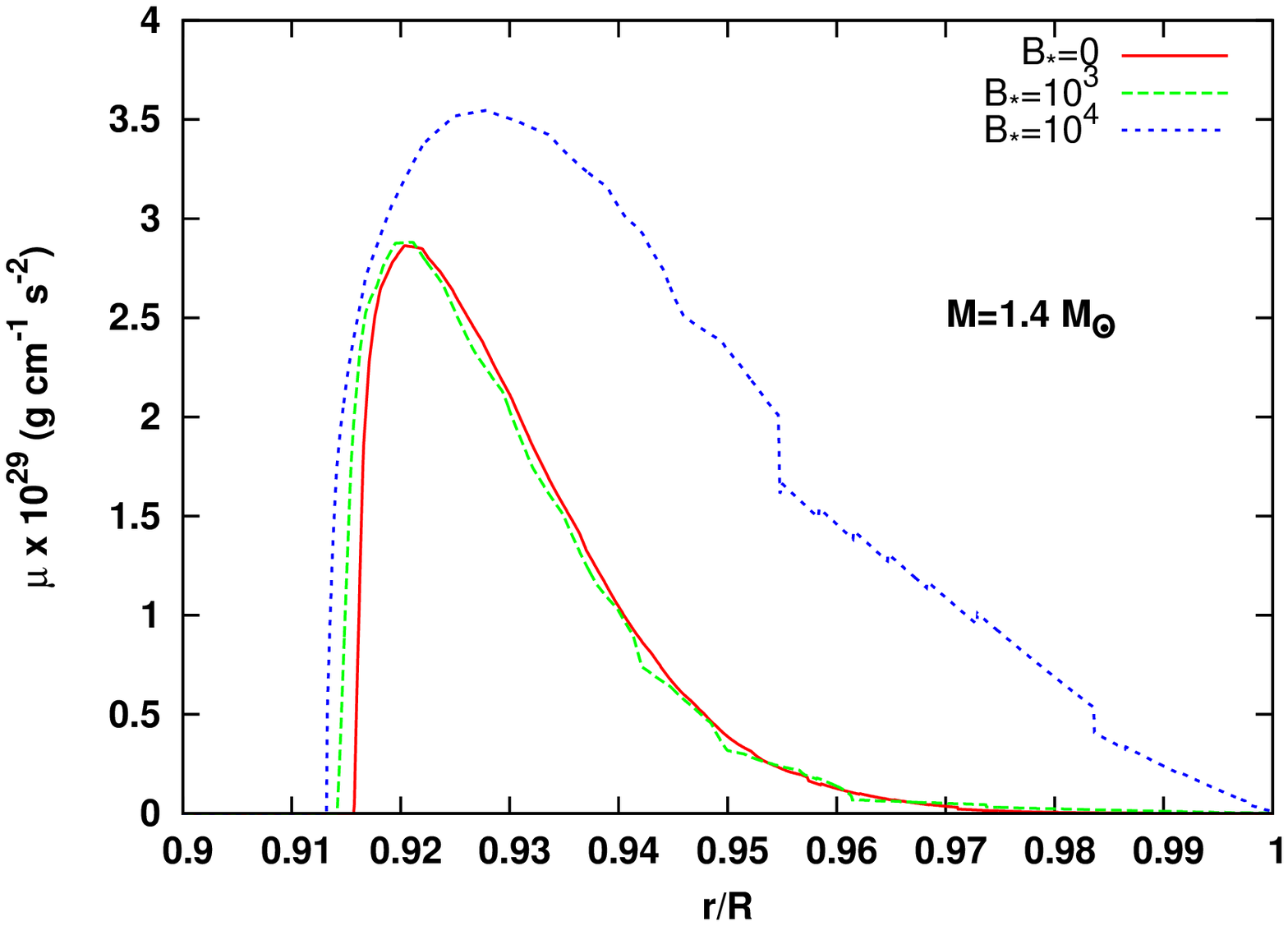}\includegraphics[width=6cm]{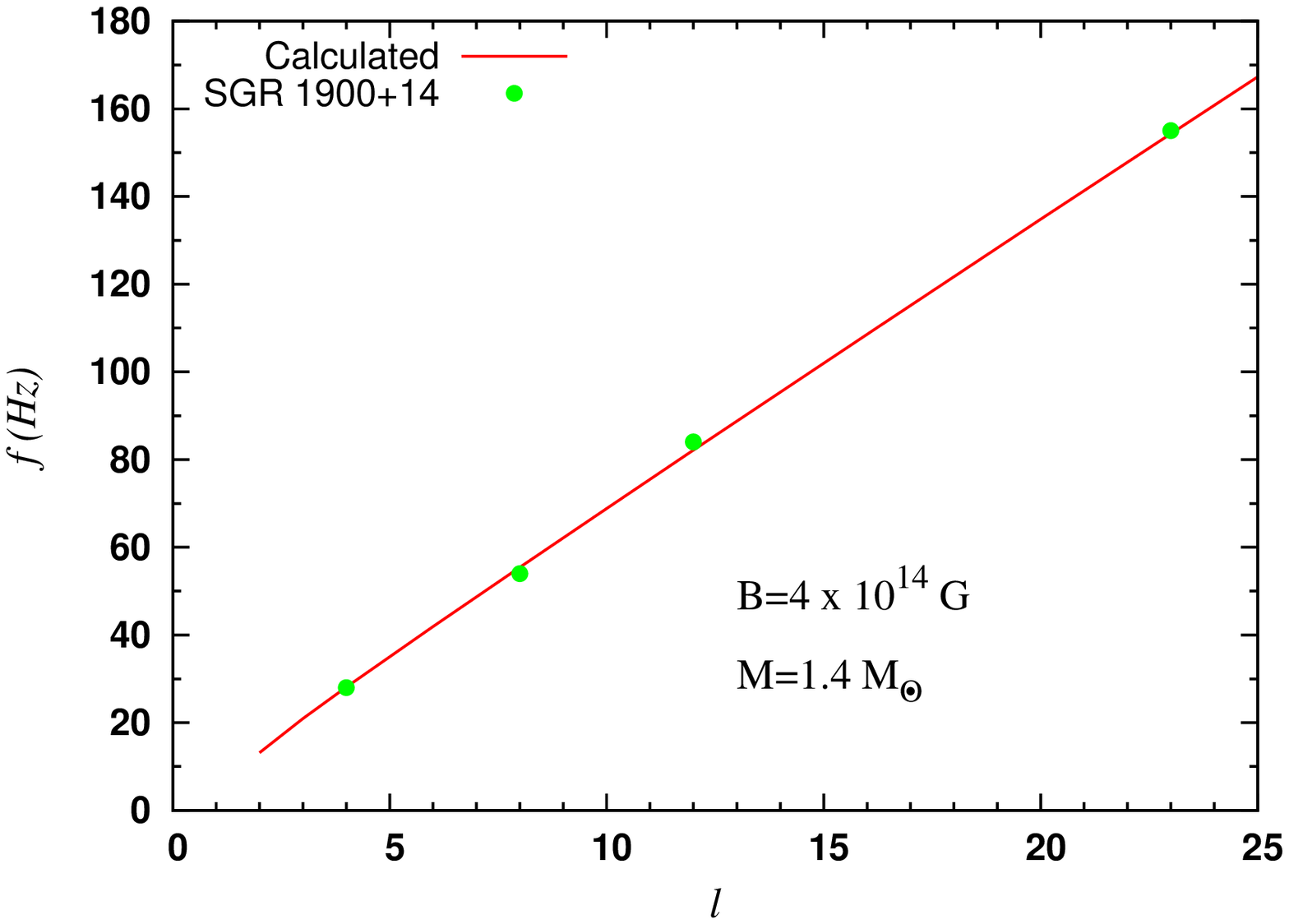}}
\caption{Shear modulus is plotted as a function of normalised distance for 
different magnetic field strengths (left panel) and shear mode 
frequencies are plotted with different $\ell$ values for a neutron star mass of
1.4 $M_{\odot}$ and $B= 4 \times 10^{14}$ G (right panel).}
\label{fig3}       
\end{figure}

We calculate the shear modulus using Eq.(\ref{shr}) and the 
compositions and equations of state of magnetised crusts obtained in Sec. 3 and
4. This is shown as a function of normalised distance with respect to radius 
($R$) of the star for different field strengths $B=0$, $B_{*}=10^3$ and 
$B_{*}=10^4$ and a neutron star mass of 1.4 $M_{\odot}$ in the left panel of 
Fig.{\ref{fig3}}. Shear 
modulus increases initially with decreasing distance and drops to zero at the 
crust-core boundary. For $B_{*}=10^4$ i.e.  
4.414$\times 10^{17}$ G or more, the shear modulus is enhanced appreciably 
compared with the zero field case.   

It was argued that shear mode frequencies are sensitive to shear modulus 
\cite{watt1,stei}. We perform our calculation for shear mode frequencies using
the model of Sotani et al. \cite{sota1} and the shear modulus of magnetised 
crusts as described above. We calculate fundamental shear mode frequencies for
a neutron star mass of 1.4 M$_{\odot}$ as well as magnetic 
fields as high as $4.414 \times 10^{17}$ G.
When we compare those frequencies involving magnetised crust with
those of the non-magnetised crust, we do not find any noticeable change between 
two cases. For SGR 1900+14 having $B=4\times 10^{14}$ G and a neutron star mass
of 1.4 M$_{\odot}$, we show in the right panel of Fig.{\ref{fig3}} that 
the observed QPO frequencies match nicely with frequencies estimated using our 
magnetised crust model. Further we observe that the first radial overtones 
calculated with our 
magnetised crust model have higher frequencies than those calculated with the 
non-magnetised crust model. This is in agreement with the prediction that
the radial overtones are susceptible to magnetic effects \cite{piro}.   

\section{Summary}
We have constructed the model of magnetised neutron star crusts and applied it
to shear mode oscillations of magnetars. In particular, we highlighted the 
effects of strongly quantising magnetic fields on the properties of ground 
state matter of outer and inner crusts in this article. It is noted that 
compositions and equations of state of neutron star crusts are significantly
altered in strong magnetic fields. Consequently, shear modulus of the crust
which is sensitive to the compositions of crusts, is enhanced. We have observed
that our model of the magnetised crust might explain the observed shear mode 
frequencies quite well.

\begin{acknowledgement}

We thank S. K. Samaddar, J. N. De, B. Agrawal, D. Chatterjee, I. N. 
Mishustin and W. Greiner for many fruitful discussions. We also acknowledge the 
support under the Research Group Linkage Programme of Alexander von Humboldt 
Foundation.
\end{acknowledgement}

\end{document}